\documentclass{aastex}%
\usepackage{emulateapj5}
\usepackage{epsfig}
\begin{document}

\slugcomment{Submitted to the Astrophysical Journal}

\def\visuncal{{\tilde{V}}}
\def\vis{V}
\def\polres{{Paper V}}
\def\paperone{{Paper I}}
\def\papertwo{{Paper II~}}
\def\check#1{{\bf #1}}
\def\stokes#1{S_#1}
\def\deg{^\circ}
\def\muK{~\mu{\rm K}}
\def\ngc{NGC~6334}
\def\rotm{{\rm RM}}
\def\snr{{s/n}}
\def\ra{{\rm R.A. }}
\def\dec{{\rm Dec.}}
\def\ran#1#2{{#1^{\rm h}}{#2^{\rm m}}}
\def\dash{ -- }

\title{Measuring Polarization with DASI} 

\author{E.\ M.\ Leitch, J.\ M.\ Kovac, C.\ Pryke, B.\ Reddall, E.\ S.\ Sandberg, M. Dragovan\footnotemark[1] and J.\ E.\ Carlstrom}
\affil{University of Chicago, 
Department of Astronomy \& Astrophysics, 
Department of Physics, 
Enrico Fermi Institute, 
5640 South Ellis Avenue, 
Chicago, IL 60637}

\smallskip
\author{and}

\author{N.\ W.\ Halverson and  W.\ L.\ Holzapfel}
\affil{University of California,
Department of Physics, Le Conte Hall,
Berkeley, CA 94720}

\begin{abstract}

We describe an experiment to measure the polarization of the Cosmic
Microwave Background (CMB) with the Degree Angular Scale
Interferometer (DASI), a compact microwave interferometer optimized to
detect CMB anisotropy at multipoles $l \simeq $~140\dash900.  The
telescope has operated at the Amundsen-Scott South Pole research
station since 2000 January.  The telescope was retrofit as a
polarimeter during the 2000\dash2001 austral summer, and throughout the
2001 and 2002 austral winters has made observations of the CMB with
sensitivity to all four Stokes parameters.  The telescope performance
has been extensively characterized through observations of artificial
sources, the Moon, and polarized and unpolarized Galactic sources.  In
271 days of observation, DASI has differenced the CMB fluctuations in
two fields to an rms noise level of $2.8\muK$.

\end{abstract}

\section{Introduction}

\footnotetext[1]{Current address: Jet Propulsion Laboratory, California Institute of Technology, 4800 Oak Grove Drive, Pasadena, CA 91109}

The DASI experiment, previously described in~\markcite{leitch02a}Leitch {et~al.} (2002)
(hereafter Paper I), is an interferometric array designed to measure
anisotropy in the cosmic microwave background radiation.  The
telescope was deployed to the Amundsen-Scott South Pole Station in the
1999\dash2000 austral summer and made total intensity measurements of the
CMB during the 2000 austral winter.  These observations were described
in Paper I. The angular power spectrum of the CMB derived from these
data were reported by \markcite{halverson02}{Halverson} {et~al.} (2002) (hereafter Paper II) and the
constraints on cosmological parameters derived from the power spectrum
were presented by \markcite{pryke02}{Pryke} {et~al.} (2002) (hereafter Paper III).  

During the 2000\dash2001 austral summer, the DASI receivers were each
fitted with broadband achromatic polarizers to allow
polarization-sensitive observations of the CMB.  In addition, a large
reflecting ground screen was installed to reduce the sensitivity to
terrestrial sources of emission. Throughout the 2001 and 2002 austral
winters, the telescope has observed the CMB in all four Stokes
parameters.  This paper (Paper IV in the continuing series) describes
the design of the DASI CMB polarization experiment, the polarization
response and calibration of the instrument, and the CMB observations
made during the 2001 and 2002 seasons. The analysis of the CMB
polarization data obtained is presented by \markcite{kovac02}Kovac {et~al.} (2002) (hereafter
Paper V).

\section{Measuring Polarization with DASI}
\label{sec:polres}

The DASI telescope has been extensively described in Paper I.  Here we
describe only hardware that has been modified or added since the
observations discussed in that paper, and briefly recapitulate details
pertinent to understanding the polarization response of the array.

DASI consists of an array of thirteen lensed, corrugated feed horns,
20~cm in diameter, each surrounded by a corrugated shroud to suppress
cross-talk between the horns.  The array operates in 10 1-GHz bands
over the frequency range 26\dash36~GHz, with horn separations ranging
from 25.1\dash120.73~cm, sampling points in the Fourier $(u,v)$ plane
at radii of 22\dash143~$\lambda^{-1}$, or equivalently, multipoles in the
range $l\simeq$140\dash900 at the frequency extrema.

The feed horns are distributed in a three-fold symmetric pattern on a
rigid faceplate which can be rotated about its axis to observe at a
range of parallactic angles.  The entire faceplate is in turn attached
to an altitude-azimuth mount.  DASI is thus a co-planar array and the
projected baselines do not change as a source is tracked around the
sky.  Given the symmetry of the array, rotations of the faceplate
which are integral multiples of $60\deg$ produce identical
Fourier-space sampling.

\begin{figure*}[t]
\begin{center}
\epsfig{file=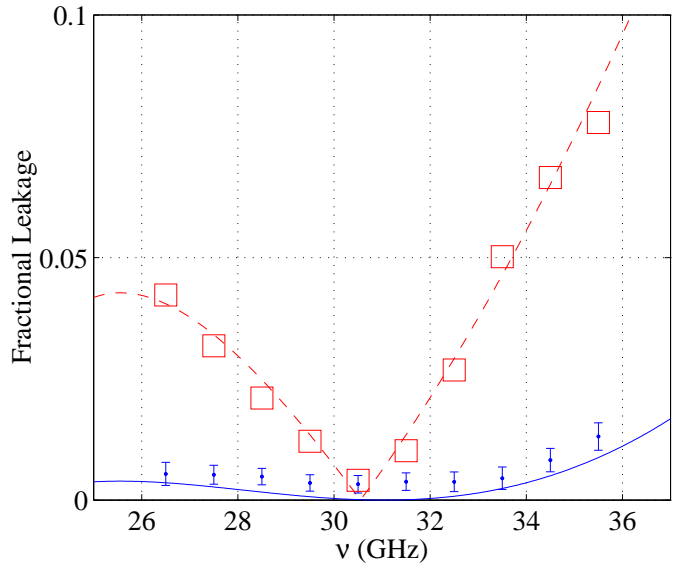,height=3in}
\end{center}
\vskip0.1in
\caption{\label{fig:polaqr} Theoretical minimum instrumental
polarization for a single-element polarizer (dashed line), and for a
2-element polarizer (solid line).  Squares are benchtop measurements
of a single-element polarizer.  Points are the astronomical
measurements of the average on-axis instrumental polarization for 
the 13 DASI receivers configured with 2-element polarizers.}
\end{figure*}

\subsection{Receiver Polarization Hardware}
\label{sec:polarizers}

The first stage of each DASI receiver is a single-moded, i.e., a
single polarization state, HEMT amplifier. DASI does not use orthomode
transformers to separate orthogonal polarization states. Instead, a
mechanically switchable waveguide polarizer is inserted between the
amplifier and the feedhorn to select between left- and right-handed
polarization states, $L$ and $R$.

The DASI horns feed circular waveguide, which by its symmetry is dual
moded. The circular waveguide is gradually tapered to single-mode
rectangular waveguide which passes only one mode of linear
polarization. For sensitivity to circular polarization, a
transformation from circular to linear polarization must therefore be
made in the circular waveguide or outside of the feed. The standard
technique uses a $\lambda/4$ retarder oriented 45$\deg$ to the
$E$-field of the linear polarization accepted by the rectangular
waveguide.  This retarder is usually a simple dielectric vane inserted
in the circular waveguide; at 30 GHz, the loss in the dielectric and
the associated noise added to the system are negligible if the vane is
cooled. The circular waveguide which holds the vane can be
mechanically rotated along the waveguide axis by $\pm 45\deg$ to
select $L$ or $R$.

DASI uses a mechanically-switched circular waveguide dielectric
polarizer similar to that outlined above.  However, a simple
$\lambda/4$ retarder leads to a contribution from the unwanted
circular polarization state, referred to as {\it leakage}, which is a
function of the offset from the design frequency.  Due to DASI's large
fractional bandwidth, the leakage for a simple $\lambda/4$ retarder
polarizer would be prohibitively large at the band edges (see
Figure~\ref{fig:polaqr}).

To construct broadband, low-leakage polarizers, we investigated
designs using multiple retarder elements and found that there are
solutions to meet any specified polarization purity and bandwidth. For
more details of the theory of multi-element achromatic waveguide
polarizers and for details of the DASI polarizers, including extensive
laboratory test results, see \markcite{kovac02b}Kovac \& Carlstrom (2002).  For the DASI polarizers
we selected a two element design composed of $\lambda/2$ and
$\lambda/4$ dielectric retarding waveguide vanes made from 1.3~mm
thick polystyrene.  With a two element design, the retarders can be
arranged so that the offset-frequency errors cancel to first
order. The design also incorporates an absorbing vane to suppress the
linear polarized mode that is reflected by the circular to
rectangular waveguide transition.

To aid in understanding the DASI design, consider the signal path as
if the receivers were actually transmitters starting with linear
polarization in the rectangular waveguide which the polarizer then
transforms to circular polarization.  In the DASI design, the
$\lambda/2$ retarder then simply introduces a rotation of the
$E$-field of the linear polarization at the design frequency, but also
introduces an offset frequency error due to the dispersion across the
band.  The $\lambda/4$ retarder is oriented $45\deg$ to the rotated
$E$-field to produce the desired circular polarization. The
orientation of the $\lambda/2$ retarder to the incident polarization
is chosen such that dispersion effects of the two retarders cancel to
first order. The entire assembly is then mechanically rotated to
select $L$ or $R$ polarization.  The offset angle between the
$\lambda/2$ and $\lambda/4$ retarders is held fixed at $60\deg$. The
angles of the $\lambda/2$ retarder to the $E$-field of the linear
polarization to produce the orthogonal circular polarization states
are $15\deg$ and $105\deg$. In Figure~\ref{fig:polaqr}, we show the
predicted leakage for the DASI 2-element design and, for comparison,
the traditional single $\lambda/4$ design.  Also shown in
Figure~\ref{fig:polaqr} is the leakage response determined from
astronomical observations, as discussed in \S\ref{sec:onaxisleak}.

The stability of the polarization leakages is more critical than their
magnitude since the leakages, if stable, can be corrected for in the
data analysis. For DASI the stability of the polarizers is set by the
repeatability and accuracy which with the angle is set.  The
polarizers are driven by a stepper motor with a $0\fdg9$ half-step,
geared down by a factor of three using anti-backlash gears, to provide
$0\fdg30$ steps of the vane assembly. The position of the polarizer is
read by an encoder with $0\fdg12$ steps. All of the hardware is
contained under vacuum within the receiver cryostats. The motor and
encoder are warm, but the waveguide vane assembly is cooled to
$\sim 10$~K.  With 2.5 counts of the polarizer encoder for the minimum
step of the motor, we can ensure a repeatable return to the same
position. We have verified that the stepper motor holds the polarizer
position to better than half of an encoder count ($<0\fdg06$) for the
entire season. This accuracy translates to a variation in the leakage
of less than 0.1\%. Astronomical tests of the stability of the
leakages on long time scales are given in \S\ref{sec:onaxisleak}.

\begin{figure*}
\begin{center}
\epsfig{file=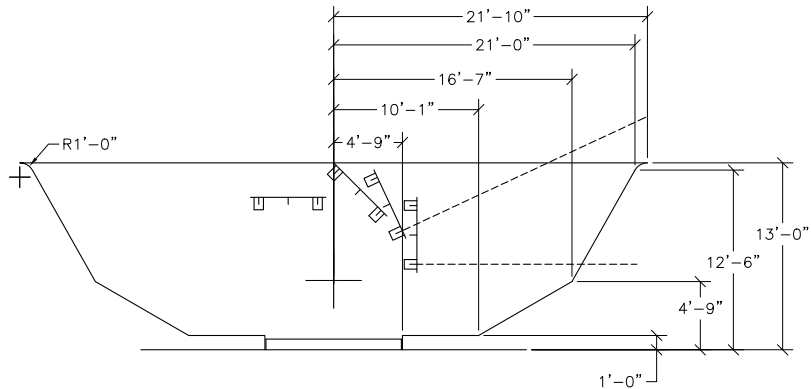,height=2in}
\end{center}
\vskip0.4in
\caption{\label{fig:dasishields} 
Ground shield geometry. The ground shields are designed to reflect 
sidelobes of the horns to the cold sky, and to prevent sidelobes
from seeing other horns, due to concerns about crosstalk.}
\end{figure*}

\subsection{Ground Shield}

A ground shield was installed around the telescope during the
2000\dash2001 austral summer, and was in place for the entirety of the
2001\dash2002 observations presented here.  The ground shield is
designed to reflect the antenna sidelobes onto the cold sky, and
consists of three concentric rings of twelve panels, the innermost
flat, the outer two rising at a pitch which increases from $\sim
36\deg$ for the inner ring to $\sim60\deg$ for the outer (see
Figure~\ref{fig:dasishields}).  The design of the panels was guided by
concerns about crosstalk between close-packed array elements; panels
are configured to prevent sidelobes up to 90$\deg$ off-axis from
seeing any other horns in reflection.

The top of the outermost ring extends to a height of $13$~feet above
the roofline.  The uppermost border of the shield is rolled with a
1~foot radius to reduce diffraction around the edge, and all gaps
between panels are covered with aluminum adhesive tape.  Four of the
upper panels are hinged and may be lowered to allow observations of
planets, or a test transmitter mounted on the roof of one of the
station buildings.  Without lowering the shields, the minimum
observing elevation is $\sim 25\deg$.

Although the DASI telescope was intended to operate with the ground
shield from its inception, the shield panels were not in place during
the year-2000 observations reported in Papers I and II, and
considerable effort was devoted to reducing the effect of the residual
ground signal in that data set.  Throughout those observations, the
effect of the ground was easily visible in the raw data and is a strong
function of $(u,v)$ radius.

During both 2000 and 2001, fields were observed in consecutive 1-hour
blocks over the same azimuth range (see \S\ref{sec:obs}), and we can
probe for residual contamination from the ground by forming the
difference between successive 1-hour means.  In the 2000 data, taken
without the ground shield, comparison of the variance in these 2-hour
differences to the variance in data differenced at the 8-second
integration rate indicates a significant ($5\%$) noise excess,
dominated by the data on the shortest baselines.  In the 2001 co-polar
data, taken with the ground shield in place, the noise excess has
dropped by nearly an order of magnitude.  Any noise excess in the
cross-polar data is a factor of 3\dash4 lower still, indicating that any
residual signal from the ground is largely unpolarized.

\subsection{Sun Shield}

When the Sun is above the horizon, data taken over much of the sky is
contaminated by signal entering the far-off sidelobes of the horn beam
pattern.  This is most evident as long-period fringing in visibility
data taken within $\sim60\deg$ of the Sun (for a definition of
visibility, see \S\ref{sec:polresponse}).  Although sunset and sunrise
at the pole occur at the March and September equinoxes, the station is
typically closed due to the excessive cold from the middle of February
until the last week in October, resulting in approximately three
months of time during which the telescope is inaccessible but the data
are unusable for CMB observations.

To extend the effective observing season, a sunshield was constructed
during the austral summer of 2001\dash2002, and was in place throughout
the 2002 observations.  The lightweight conical shield, constructed
from the top $\sim3$ feet of a spare shield from the TopHat
experiment, attaches rigidly to the front face of the telescope,
completely enclosing, but not rotating with, the faceplate.  With a
diameter at the attachment point of $62.93''$, and an opening angle of
$15\deg$, the shield prevents any sidelobe of an antenna from seeing
the Sun when the telescope is pointed $> 65\deg$ away.

Contamination from the Sun was plainly evident in the 2000 data as
fringing in the calibrator visibility data taken after sunrise, and is
evident as increased scatter on the shortest baselines even in the
2001 data taken with the ground shield present.  These data show
excursions in the visibility amplitude on the shortest baselines of
$\sim50\%$ about the pre-sunrise means, although the longest baselines
are unaffected.  In data taken during the 2002 sunset with the
sunshield present, there is still some evidence for increased scatter
in the short-baseline visibilities, but the effect is reduced by
nearly two orders of magnitude.  For fields $>90\deg$ from the Sun,
the visibility scatter is consistent with the noise on all baseline
lengths.

The sunshield was left in place throughout the 2002 winter season,
and it is found that during periods when the Moon is above the
horizon but below $10\deg$ elevation, the data from 2002 show evidence
of excess noise, while the 2001 data do not.  We speculate that during
periods when the Moon is within the opening angle of the sunshield,
secondary reflection off the shield can actually contaminate the
short-baseline data when the Moon would otherwise be screened by the
ground shield.  Accordingly, a stricter cut on the Moon elevation is
applied to the 2002 data than to the 2001 data (see
\S\ref{sec:calcuts}).

\subsection{Polarization Response}
\label{sec:polresponse}

To derive the polarization response of an interferometer, consider the
complex electric field incident on a single receiver, ${E}_{m}$.  The
response of the interferometer on a baseline $m$--$n$, called the {\it
visibility}, is simply the time-averaged cross-correlation of the
fields from a pair of receivers $(m,n)$:

\begin{equation}
\vis_{mn}=\langle{{E}_{m}{E}^*_{n}}\rangle.
\end{equation}
Rewriting ${E}_{m}$ in terms of its components in an arbitrary
coordinate system rotated by an angle $\psi$ relative to a coordinate
system fixed to the receivers (i.e., parallactic angle), the usual
Stokes parameters can be substituted to obtain

\begin{eqnarray}
\nonumber{\vis^{RR}_{mn}} &=& {1\over{2}}\left\{\stokes{I}+\stokes{V}\right\}\\\nonumber
{\vis^{LL}_{mn}} &=& {1\over{2}}\left\{\stokes{I}-\stokes{V}\right\}\\\nonumber
{\vis^{RL}_{mn}} &=& {1\over{2}}\left\{\stokes{Q}+i\stokes{U}\right\}e^{-2i\psi}\\
{\vis^{LR}_{mn}} &=& {1\over{2}}\left\{\stokes{Q}-i\stokes{U}\right\}e^{2i\psi}
\label{eq:vis}
\end{eqnarray}
for receivers set to receive right circular ($R$) and left circular
($L$) polarization.  We refer to these four combinations as {\it
Stokes states}, by analogy with the Stokes parameters which can be
derived from them.

Since the DASI correlator was originally designed to accommodate only a
single Stokes state per IF, observation in all four states is achieved
via time-multiplexing.  The polarizer for each receiver is switched
between $L$ and $R$ on a Walsh sequence which spends an equal amount
of time in each state, i.e., $\sum_{i}{f_i} = 0$ if we represent
the two states as $f \in \pm1$.  The product of two Walsh functions is
another Walsh function, whence $\sum_{i,n\neq m}{f_i^nf_i^m} = 0$,
and the baseline therefore spends an equal amount of time in the
co-polar ($RR$, $LL$) and cross-polar ($LR$, $RL$) states.  In fact, it
can be shown that with this sequencing of the polarizers, over the
course of a full Walsh cycle each baseline of the interferometer
spends an equal amount of time in each of the four Stokes states,
although different baselines will sample different Stokes states at
any given time.  For the observations presented here, a Walsh function
of period 16 with a time step of 200 seconds was used, so that over
the course of an hour, every Stokes state is sampled by every baseline
for approximately 13 minutes.

\subsection{Relative Gain and Phase Calibration}
\label{sec:calib}

In practice, each receiver introduces a unique phase offset and
amplitude variation to the electric field, so that the expression for
the detected field must be modified by a receiver-dependent complex
gain, $\tilde{E}_m = g_mE_m$, where $g_m$ depends on the polarization
state of the receiver.  Furthermore, the correlator can introduce
baseline-dependent complex gains $g_{mn}$, independent of polarization
state, so that the measured visibilities are given by:

\begin{eqnarray}
\nonumber
{\visuncal^{RR}_{mn}} &=& g^R_m{g^R_n}^*g_{mn}{\vis^{RR}_{mn}}\equiv G^{RR}_{mn}{\vis^{RR}_{mn}}\\\nonumber
{\visuncal^{LL}_{mn}} &=& g^L_m{g^L_n}^*g_{mn}{\vis^{LL}_{mn}}\equiv G^{LL}_{mn}{\vis^{LL}_{mn}}\\\nonumber
{\visuncal^{RL}_{mn}} &=& g^R_m{g^L_n}^*g_{mn}{\vis^{RL}_{mn}}\equiv G^{RL}_{mn}{\vis^{RL}_{mn}}\\
{\visuncal^{LR}_{mn}} &=& g^L_m{g^R_n}^*g_{mn}{\vis^{LR}_{mn}}\equiv G^{LR}_{mn}{\vis^{LR}_{mn}}.
\label{eq:measvis}
\end{eqnarray}

Since the co-polar visibilities ${\visuncal^{RR}_{mn}}$ and
${\visuncal^{LL}_{mn}}$ are just proportional to Stokes $\stokes{I}\pm
\stokes{V}$, the gain factors $G^{RR}_{mn}$ and $G^{LL}_{mn}$ can
simply be derived on a per-baseline basis from observations of a
bright unpolarized point source, or more complicated source whose
intrinsic visibility structure is known (if we take the circular
polarization $\stokes{V}$ to be zero).

However, we require a source of known linear polarization to derive
the complex gains in the same manner for the cross-polar visibilities
${\visuncal^{RL}_{mn}}$ and ${\visuncal^{LR}_{mn}}$.  These are in
general scarce at high frequencies, and furthermore few are well
studied in the southern hemisphere.  The low gain of DASI's
20-cm feed horns, moreover, makes it impracticable to observe all but
the brightest sources, in total intensity or otherwise, and the
brightest sources at $30$~GHz are typically compact HII regions, whose
polarization fraction is expected to be negligible.

\begin{figure*}[t]
\begin{center}
\epsfig{file=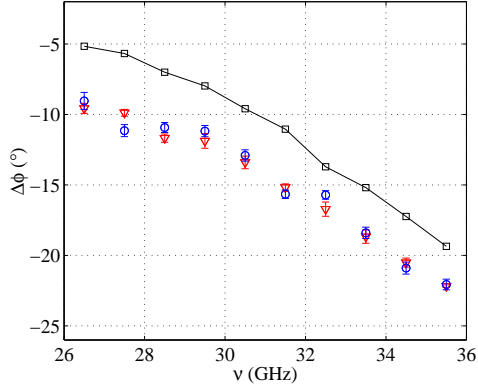,width=2.5in}
\end{center}
\caption{The absolute phase offsets determined from wire grid
observations (triangles) and from the Moon (circles).  Shown also are
benchtop measurements of the same polarizer (solid line, squares),
with an arbitrary offset subtracted.}
\label{fig:absphase}
\end{figure*}

As can be seen from Equation~\ref{eq:measvis}, however, we can in
principle derive the cross-polar gains from co-polar observations of
an unpolarized source, if we work in ratios of complex gains:
\begin{equation}
{\visuncal^{LL}_{mn}\over{\visuncal^{RR}_{mn}}} =
\left({g^L\over{g^R}}\right)_m{\left({g^L\over{g^R}}\right)^*_n}
\equiv r_m{r^*_n}
\end{equation}
in which the correlator gains $g_{mn}$ exactly cancel.  For DASI, this
yields an overconstrained set of 78 complex equations which can be
solved for the 13 antenna-based ratios $r_k = g^L_k/g^R_k$.

The solution is not unique, and we must arbitrarily specify the phase
$\Delta \phi \equiv \phi_R-\phi_L$ of one of the ratios, typically
setting it to zero.  With the ratios $r_k$ in hand, the baseline
complex gains for the cross-polarized channels can be recovered from
the co-polar solutions as:
\begin{eqnarray}
\nonumber
r^*_nG^{RR}_{mn} &=& {\left({g^L\over{g^R}}\right)^*_n}g^R_m{g^R_n}^*g_{mn} = 
e^{-i\Delta\phi}G^{RL}_{mn} \\
r_mG^{RR}_{mn} &=& \left({g^L\over{g^R}}\right)_mg^R_m{g^R_n}^*g_{mn} = 
e^{i\Delta\phi}G^{LR}_{mn} 
\end{eqnarray}
that is, we can construct the cross-mode calibration factors from the
antenna-based solutions, but only up to the unknown phase difference
between $R$ and $L$ for our reference antenna.  This overall phase
offset, while leaving the amplitude of the cross-polar visibilities
unaffected, will mix power between $\stokes{Q}$ and $\stokes{U}$, or
between $E$ and $B$, in terms of the decomposition into the
coordinate-independent scalar fields which has become the standard in
the CMB polarization literature \markcite{kamionkowski97,zaldarriaga97,hu_w97}(Kamionkowski, Kosowsky, \&  Stebbins 1997; {Zaldarriaga} \& {Seljak} 1997; Hu \& White 1997).  Removal of the offset is therefore
critical to obtaining a clean separation of CMB power into $E$- and
$B$-modes.

In practice, we use the bright, compact, unpolarized HII region,
RCW38, to determine the co-polar and cross-polar calibration factors
in this manner.  The phase offset can only be determined by observing
a source whose plane of polarization is known, as described in the
next section.

\subsection{Absolute Phase Calibration}
\label{sec:absphase}

To determine the phase offset introduced by the complex gain
correction described above, we require a source whose polarization
angle (though not amplitude) is known.  As discussed in
\S\ref{sec:calib}, however, few suitably strong polarized sources are
available, and in practice we create a source of known polarization
angle by observing an unpolarized source through polarizing wire
grids.  At centimeter wavelengths, the required wire spacing
($\lesssim \lambda/3$) poses no particular technical challenges, and
grids can be simply constructed out of conventional materials;
thirteen wire grids were constructed out of 34-gauge copper wire,
wound with a spacing of $0.05''$.  These grids attach directly to the
exterior of the corrugated shrouds and completely cover the aperture
of the DASI horns.

The same source used for gain calibration of the visibilities, RCW38,
was observed with the wire grids attached in two orientations, one
with wires parallel to the ground, and the other with grids (but not
receivers) rotated by 45 degrees with respect to the first.  It was
confirmed that in the second orientation the apparent phase of the
cross-polar visibilities shifted by $\pm 90\deg$ relative to the
first, as indicated by Equation~\ref{eq:vis}.

The measured absolute phase offsets $\Delta\phi(\nu)$ are shown in
Figure~\ref{fig:absphase}.  These offsets show a trend in frequency
which is a direct consequence of the achromatic polarizer design
discussed in \S\ref{sec:polarizers}.  As Figure~\ref{fig:absphase}
shows, this trend is in excellent agreement with expectations from
benchtop measurements of the polarizers, where the expected trend is
shown with an arbitrary offset subtracted.  These phase offsets were
measured in 2001 August and again in 2002 February and were found to
agree within measurement errors.  As discussed below, we can restrict
the intrinsic polarization fraction of RCW38 to $< 0.09\%$ at our
observing frequency; since the grids create a polarized source with an
effective polarization fraction of $100\%$, any intrinsic polarized
flux at this level will have a negligible effect on the measurements
of the absolute phase offsets.

Although suitably bright point sources of known polarization angle are
not available, we can in principle derive the absolute phase offsets from
observations of the Moon, as discussed in \S\ref{sec:moon}.  These
results, also shown in Figure~\ref{fig:absphase}, are consistent
with the offsets measured with the wire grid polarizers, to within the
$\sim0\fdg4$ measurement errors.  As discussed in \polres, errors in
the absolute phase offset of this magnitude have a negligible effect
on the power spectral analysis.

\begin{figure*}[t]
\epsscale{1.5}
\plotone{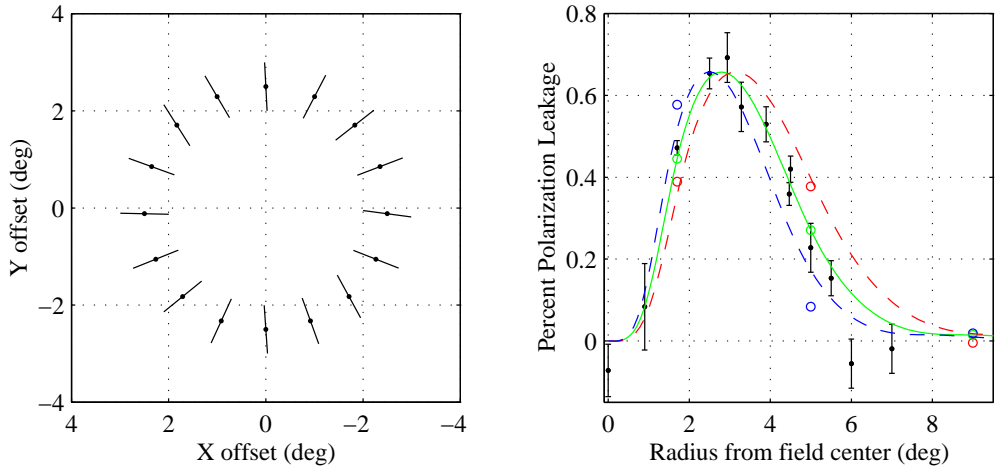}
\caption{(left) The measured direction of the instrumental leakage for
a ring of $2\fdg5$ offset pointings on RCW38.  (right) The amplitude
profile of the instrumental polarization shown versus radius from
field center.  Moon data are shown as red, green and blue circles
corresponding to the frequency bands 26\dash29, 29\dash33, and 33\dash36~GHz
respectively.  RCW38 data for the full 26\dash36~GHz band are shown as
black points.  The green line is a spline interpolation of the
29\dash33~GHz Moon data, and the red and blue dashed lines are the
interpolation scaled to 27.5 and 34.5~GHz respectively.  See text for
further discussion.  }
\label{fig:offaxisleak} 
\end{figure*}

\subsection{Instrumental Polarization}

For perfect polarizers, circular polarization can be thought of as the
superposition of two orthogonal linear modes, $E = \alpha E_x + \beta
E_y$, where one mode leads the other by exactly $90\deg$, i.e.,
$\beta/\alpha = e^{\pm i\pi/2}$.  For perfect polarizers, the
cross-polar visibilities are strictly linear combinations of the
linear polarization measures, Stokes $\stokes{Q}$ and $\stokes{U}$, as
indicated by Equation~\ref{eq:vis}.  For realistic polarizers,
however, a small amount of radiation of one handedness will be passed
by a polarizer set to transmit the opposite handedness, leading to a
term in the expression for the cross-polar visibilities proportional
to the total intensity.  In the limit of small deviations from perfect
orthogonality, $\beta/\alpha \simeq i(1 - \epsilon)$, and keeping only
terms to first order in $\epsilon$, it can be shown that the
cross-polar visibilities are given by:

\begin{eqnarray}
\nonumber
{\vis^{RL}_{mn}} &=&
{1\over{2}}\left\{(\stokes{Q}+i\stokes{U})e^{-2i\psi} + \stokes{I}(\epsilon^R_m +
{\epsilon^L_n}^*)\right\}\\
{\vis^{LR}_{mn}} &=&
{1\over{2}}\left\{(\stokes{Q}-i\stokes{U})e^{2i\psi} + \stokes{I}(\epsilon^L_m +
{\epsilon^R_n}^*)\right\}
\label{eq:leak}
\end{eqnarray}
The terms $L^{RL}\equiv\epsilon^R_m + {\epsilon^L_n}^*$ and
$L^{LR}\equiv\epsilon^L_m + {\epsilon^R_n}^*$ are sums of the
polarizer leakages discussed in \S\ref{sec:polarizers}, and are
expected from benchtop measurements of the DASI polarizers to be on
average $< 1\%$ (see Figure~\ref{fig:polaqr}).  As is obvious from
Equation~\ref{eq:leak}, the presence of leakages has the effect of mixing
$\stokes{I}$ into $\stokes{Q}$ and $\stokes{U}$, or equivalently, the
leakages will mix CMB power from $T$ into $E$ and $B$ (see also \polres).

\subsubsection{On-Axis Leakage}
\label{sec:onaxisleak}

As discussed in \S\ref{sec:polres},
the receivers are mounted on a rigid
faceplate which can be driven to simulate parallactic angle rotation.
From Equation~\ref{eq:leak}, it can be seen that the leakages are
irrotational, while any contribution to the visibilities from
intrinsic source polarization will have a phase modulated by the
faceplate rotation.  The data can therefore be fit for an offset plus
a sinusoidal modulation to isolate the leakages.  Equivalently, when
observations are made at three or six-fold symmetric faceplate
rotations, the cross-polar visibilities for a radially symmetric
source can simply be averaged over faceplate rotations to cancel the
intrinsic term.

To determine the magnitude of the leakages, RCW38 was observed at six
faceplate rotations separated by $60\deg$, for approximately 3 hours
at each faceplate position, resulting in an rms noise on the baseline
leakages of $0.8\%$ (of $\stokes{I}$).  As can be seen from
Equation~\ref{eq:leak}, however, the baseline leakages are simply linear
combinations of antenna-based terms, and can therefore be solved for
the antenna-based leakages $\epsilon$, yielding a typical error of
$0.3\%$ on the antenna-based leakages.  Typical antenna-based leakage
amplitudes range from $\lesssim 1\%$ over much of the frequency band,
to $\sim 2\%$ at the highest frequency.  The averages of the
antenna-based leakages $\epsilon^{R,L}$ are shown in
Figure~\ref{fig:polaqr}.  Prior to installation on the telescope, the
polarizers were optimized in a test setup to minimize the leakage
response; as can be seen in Figure~\ref{fig:polaqr}, these
observations indicate that we have achieved close to the theoretical
minimum.  As discussed in \polres, leakages of this magnitude have a
negligible effect on the separation of the CMB signal into $E$ and $B$
modes.

Of critical importance is the stability of the leakages, as they
cannot be frequently measured.  The leakages were measured once in
August 2001, and again in April 2002 and July 2002, and the baseline
leakages show excellent agreement between all three epochs.  The
higher \snr\ antenna-based leakages show similarly good agreement
between August 2001 and April 2002, with residuals at all frequencies
$\lesssim 1\%$, with the exception of three receivers.  The polarizers
for these receivers were retuned during the 2001\dash2002 austral
summer, and systematic offsets can clearly be seen in the
antenna-based residuals between these two epochs, the largest being
approximately $2.5\%$ in amplitude.  Residuals between the April 2002
and July 2002 leakages are $\lesssim 1\%$ at all frequencies.
Variations in the leakages of this magnitude are expected to have a
negligible impact on the analysis presented in \polres.

As discussed above, the parallactic angle modulation can be used to
eliminate the leakage and place a limit on the intrinsic source
polarization.  Averaging over the three epochs at which the leakages
were measured, it is found that the polarization amplitude of RCW38,
$P = (\stokes{Q}^2 + \stokes{U}^2)^{1/2}$, is less than $0.09\%$ (of
$\stokes{I}$) at all frequencies.

\begin{figure*}[t]
\begin{center}
\epsfig{file=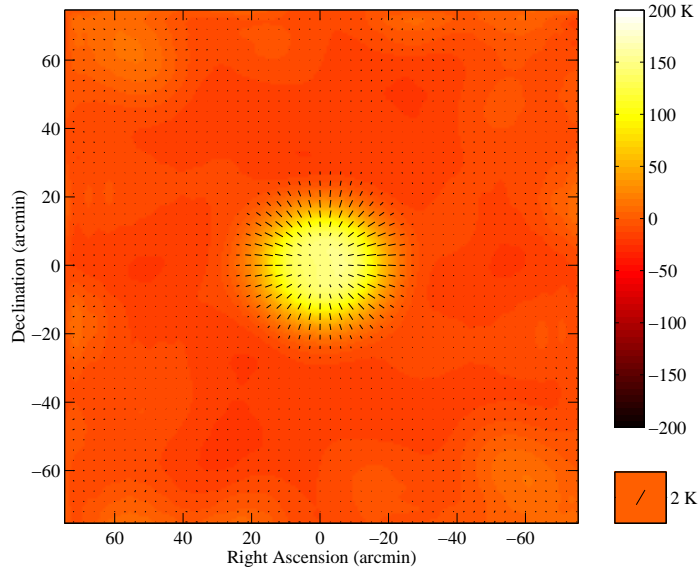,height=3in}
\end{center}
\caption{\label{fig:moon} DASI image of the Moon.  Shown is the total
intensity map (greyscale), with measured polarization vectors
overplotted, demonstrating the expected radial polarization pattern.
These maps are dirty maps (see text) with $\sim 22'$ resolution, and
the conversion to temperature units is therefore only approximate.
}
\end{figure*}

\subsubsection{Off-Axis Leakage}
\label{sec:offaxisleak}

Although the polarizers were optimized to yield low leakage, the lensed feed
horns
themselves will induce an instrumental polarization that varies across
the primary beam. The instrumental polarization of the circularly 
symmetric lenses and feeds is expected to be azimuthally symmetric and 
depend only on the angle from the phase center.

To probe the off-axis response of the feeds, observations of RCW38
were made with the source at 16 offset positions in a $2\fdg5$ radius
ring.  The direction of the instrumental leakage is shown in the left
panel of Figure~\ref{fig:offaxisleak}.  Within the measurement
uncertainty this data is consistent with a simple radial pattern.  In
terms of $\stokes{Q}$ and $\stokes{U}$, a radial pattern produces a
quadrupole asymmetry across the primary beam, with the $\stokes{Q}$
and $\stokes{U}$ patterns rotated by $45\deg$ with respect to one
another.

The radial amplitude profile of this off-axis leakage was
characterized by observations of RCW38 and the Moon at various offsets
(see Figure~\ref{fig:offaxisleak}). The Moon data have extremely high
\snr, permitting a separation of the data into several frequency
bands.  Attempts to model the off-axis leakage data as a simple
perturbation of the aperture field reproduce the general shape and
frequency dependence of the measured profile, but fail to replicate
the sharp rise in amplitude seen between $1\deg$ and $2\deg$.  We
therefore fit an empirical model by spline interpolation of the Moon
data for the 29\dash33~GHz band (together with a point from RCW38 at
$0\fdg9$ offset).  The effect is assumed to scale with frequency in
the same manner as the aperture field, and indeed this is supported by
the Moon data at other frequencies.  Additionally, the beam-offset
RCW38 data confirms the validity of the model, agreeing within the
measurement uncertainty.

The presence of off-axis leakage will transform CMB power from
$\stokes{I}$ into $\stokes{Q}$ and $\stokes{U}$, and will bias the
estimation of polarization parameters from the data.  These effects
were investigated via extensive Monte Carlo simulation using the
off-axis leakage model shown in Figure~\ref{fig:offaxisleak}.  As is
shown in \polres, the resulting bias on the various likelihood results
presented in that paper is $\lesssim 4\%$ if the off-axis leakage is
not accounted for in the analysis. Including our model of the off-axis
leakage in the analysis, we can eliminate the bias to 1\% or better.

Off-axis leakage will also transform power from point sources in the
anisotropy fields into a polarization signal. In \polres, we discuss
the results of extensive simulations of this effect using the best
available point source counts at cm-wavelengths. The results indicate
that the bias on the derived CMB parameters is again small, of order a few
percent or less.

\section{Polarization observations with DASI}

During the course of the 2001\dash2002 CMB observations, considerable
effort was expended to characterize the polarization response of the
instrument through observations of astronomical sources.  In the
preceding sections, we have described at length observations of the
unpolarized HII region RCW38, on which the bulk of DASI's calibration
is based. Here we present polarimetric observations of a variety of
other sources which provide independent checks of those calibrations,
and lend confidence that we understand the various aspects of the
instrumental response to a level (precision better than 1\%) beyond that
required to measure polarization in the CMB.

\subsection{Observations of the Moon}
\label{sec:moon}

At centimeter wavelengths, radiation from the Moon is dominated by
thermal emission from the regolith, typically from the first
$\sim10$\dash20~cm of the Moon's surface.  This radiation is intrinsically
unpolarized, but scattering off the dielectric discontinuity at the
surface will induce polarization; the tangentially polarized component
is preferentially reflected, leading to a net radial polarization
pattern across the disk of the Moon.  This polarization amplitude
decreases to zero at the center and increases to a maximum at the limb
(see \markcite{moffat72}{Moffat} (1972) for a comparable result at 21 cm, and
\markcite{mitchell94}{Mitchell} \& {De Pater} (1994) for a similar discussion and map of the microwave
polarization of Mercury).

Observations of the Moon were made with DASI during 2001\dash2002 at
several epochs, and at various phases of the Moon.  
All show excellent agreement with the expected radial polarization
pattern, and the consistency between epochs attests to the stability
of the absolute phase offset.  In Figure~\ref{fig:moon} we present a
polarized map of the Moon made during 21\dash22 August 2001, when the
Moon was nearly full.  These data have been corrected for instrumental
leakages and the absolute phase offset determined from wire grid
observations, discussed in \S\ref{sec:absphase}.

Note that any residual phase offset in the cross-polar visibilities
will introduce a vorticity to the polarization vectors, from which we
can derive an independent measure of the cross-polar phase offsets.
These measurements are shown in Figure~\ref{fig:absphase}, and are in
excellent agreement with the phase offsets determined from the wire
grid observations discussed in \S\ref{sec:absphase}.

\begin{figure*}[t]
\epsscale{2}
\plotone{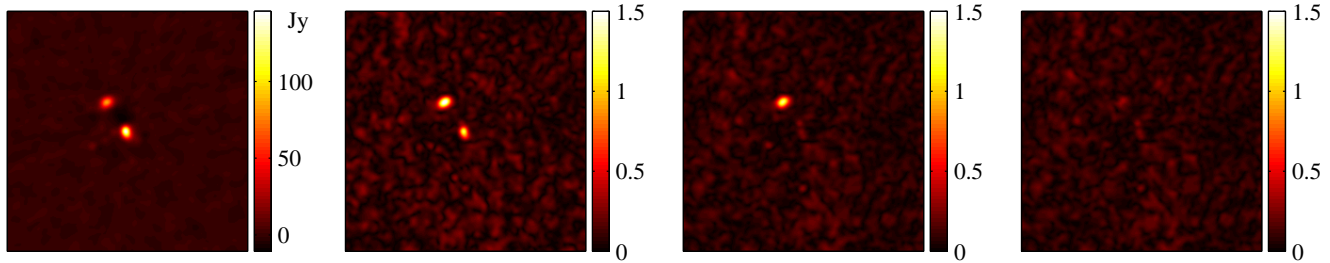}
\caption{Observations of the molecular cloud complex \ngc.  (left
panel) Total intensity map. (center left) Map of $P = ({\stokes{Q}^2 +
\stokes{U}^2})^{1/2}$, uncorrected for leakage, showing structure
correlated with the total intensity map, and at a level consistent
with instrumental polarization. (center right) Map of $P$ corrected
for on-axis leakage.  The central source has disappeared, and the
on-axis residuals are consistent with intrinsic source polarization
$<0.14\%$.  (right panel) Map of $P$, corrected for off-axis leakage,
using the profiles shown in Figure~\ref{fig:offaxisleak}.  Fields are
$12\fdg8$ across, with R.A. increasing to the left.  All intensity
scales are in Jy.}
\label{fig:ngc6334} 
\end{figure*}

\subsection{Observations of Galactic Sources}

\subsubsection{\ngc}
\label{sec:ngc6334}

Polarimetric observations of the Galactic source \ngc~were made in
August 2001, centered on $\ra = \ran{17}{20}$, $\dec = -35\deg50'$ (J2000).
\ngc~is a massive molecular ridge, with numerous embedded
photo-dissociation regions, exhibiting the rich phenomenology typical
of massive star-forming complexes.  At radio and infrared wavelengths,
emission from this complex encompasses a broad variety of phases of
the interstellar medium, ranging from line and continuum emission from
molecular gas and dust, to optically thick and dust-obscured free-free
emission from ionized gas in embedded HII regions (see for example
\markcite{burton00}{Burton} (2000), and \markcite{jackson99}{Jackson} \& {Kraemer} (1999)).

In Figure~\ref{fig:ngc6334}, we present maps of the source in total
intensity and polarization fraction.  Artifacts of the Fourier
sampling of the array have been reduced in these maps by an
application of the CLEAN algorithm \markcite{hogbom74}(H\"{o}gbom 1974). At DASI's
$\sim22'$ resolution, the microwave emission from the ridge is
concentrated in two regions of comparable flux, one on-center, the
second near the half-power point of the primary beam.  This source
therefore provides an ideal testbed for the leakage corrections
described in \S\ref{sec:onaxisleak} and \S\ref{sec:offaxisleak}.  As
can be seen in the second panel of Figure~\ref{fig:ngc6334}, the
$\stokes{Q}$ and $\stokes{U}$ maps show emission coincident with these
regions, at a level consistent with the instrumental leakages.
Application of the on-axis leakage correction to the visibilities
completely removes the central source in the polarization map,
demonstrating that we can correct for instrumental leakage to better
than $0.2\%$.  Applying the off-axis leakage profile in the image
plane similarly accounts for the second source.

The residuals after correction are consistent with the noise in the
polarization map, indicating that the net polarization fraction from
the variety of emission processes represented in the source is $<
0.14\%$.

\subsubsection{Vela}
\label{sec:vela}

As an application of the various calibrations discussed above, and a
direct demonstration of DASI's ability to map complex wide-field
polarized emission, we present observations of the Vela supernova
remnant.  Inspection of the Parkes 2.4 GHz polarization maps
\markcite{duncan97}({Duncan} {et~al.} 1997) shows that the Vela complex is an exceptional region,
both for its size and intensity, and for its high fractional
polarization ($\sim30$\%).  Observations were made of a field centered
at $\ra = \ran{08}{35}$, $\dec = -45\deg46'$ (J2000).

\begin{figure*}[t]
\epsscale{2}
\plotone{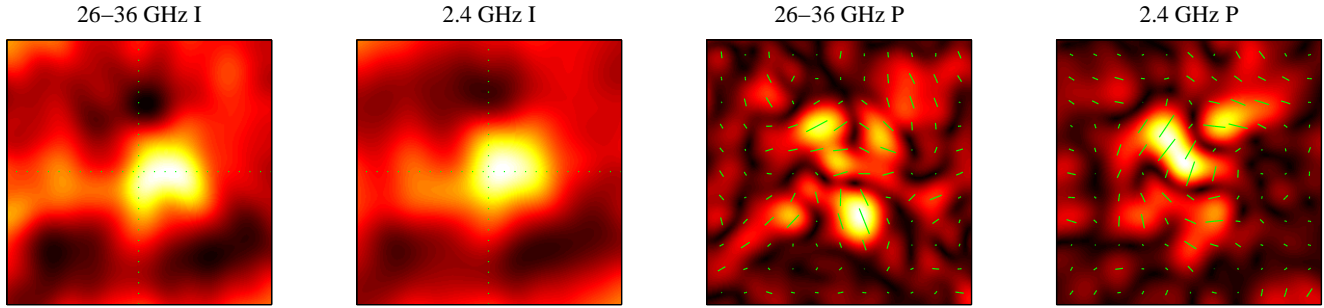}
\caption{``Dirty maps'' of the Vela region from DASI
and as constructed from the Parkes 2.4 GHz survey \markcite{duncan97}({Duncan} {et~al.} 1997)
with equivalent spatial filtering applied.
(left pair) Color scale shows total intensity for the two cases.
(right pair) Color scale shows polarized intensity ($P = ({\stokes{Q}^2 +
\stokes{U}^2})^{1/2}$) with vectors showing the electric field direction
(length of vectors is proportional to $P$).
Each map is $3\deg$ square.
}
\label{fig:velapol} 
\end{figure*}

The left and center right panels of Figure~\ref{fig:velapol} show the
resulting maps.  Artifacts of the Fourier sampling have not been
deconvolved from the map, i.e., these are ``dirty'' maps.  In
addition, we produced simulated DASI visibilities from the 2.4~GHz
maps and made dirty maps from these with the same weightings as for
the DASI data.  The two sets of maps are therefore directly
comparable, and within the limitations of noise will be the same in as
much as the distribution of 26\dash36~GHz emission tracks that at
2.4~GHz.  The 2.4~GHz maps are shown in the center left and far right
panels of Figure~\ref{fig:velapol}.

As can be seen in the polarization maps, the 2.4 GHz vectors appear
systematically rotated with respect to the 26\dash36~GHz vectors.  A
histogram of the difference between the polarization vector angles,
weighted by the polarization amplitude, shows a clear peak at +1~rad.
In the absence of a phase wrap, this implies a rotation measure of
$\sim +70$~rad~m$^{-2}$ which is of the same order of magnitude as
nearby line of sight measurements~\markcite{simard81a}({Simard-Normandin}, {Kronberg}, \&  {Button} 1981).

\section{CMB Observations}
\label{sec:obs}

\subsection{Field Selection}
\label{sec:field_select}

For the total intensity anisotropy measurements presented in
\paperone, we observed 4 rows of 8 fields separated by 1~h in right
ascension.  The locations of these fields were chosen as a compromise
between observing at high elevation to minimize ground signal, and in
regions showing very low levels of Galactic emission in both the IRAS
100 micron and Haslam 408~MHz maps \markcite{haslam81}({Haslam} {et~al.} 1981).  We were able
clearly to identify point sources from our own data within these
fields as described in \paperone.

For the polarization observations presented here, we selected the C2
and C3 fields centered at $\ra = \ran{23}{30}$ and $\ran{00}{30}$, and $\dec =
-55\deg$, within which no point sources had been detected.  These fields
lie at Galactic latitude $-58\fdg4$ and $-61\fdg9$, respectively.  The
brightness of the IRAS 100 micron and 408~MHz maps within our fields
lie at the 6\% and 25\% points respectively of the integral
distributions taken over the whole sky.

\subsection{Observations}

The fields were tracked over the full azimuth range in two-hour
blocks.  In any block, each field was tracked for approximately 1 hour
over the same azimuth range, allowing a constraint on any residual
ground signal not removed by the shields.  Observations were divided
into self-contained 24-hour segments.  Each segment comprised 20 hours
of CMB observation, and bracketing observations of the primary
calibrator source RCW38.  This source has been described at length in
\paperone, and the reader is referred to that paper for details.

During 2000, observations of RCW38 performed every 12 hours
demonstrated that instrumental gains were stable at the $\sim1\%$
level over many days; during 2001\dash2002, the source was therefore
observed only at the beginning and end of the 20-hour CMB
observations.  In each calibrator scan, the source was observed for 35
minutes in each of the $RR$ and $LL$ configurations, from which the
instrumental gains were determined for both the co-polar and
cross-polar visibilities, as discussed in \S\ref{sec:calib}.  In 35
minutes, we achieve an accuracy of $\lesssim 2\%$ on both the co-polar
and cross-polar visibilities.

These observations were bracketed by skydips from which the opacity
was determined, and by injection of a noise source used to calibrate
the complex correlator, as described in \paperone.

To provide an additional level of consistency checks, observations
were performed at two distinct rotations of the array faceplate (see
\S\ref{sec:polres}), separated by $60\deg$.  Each 24-hour segment was
observed at a fixed faceplate position, but the faceplate position was
alternated roughly every day, so that approximately half the data from
2001\dash2002 were acquired in each orientation.

Fields were observed from 2001 April 10 to 2001 October 27, and again
from 2002 February 14 to 2002 July 11.  In all, 271 days of usable CMB
data were acquired, 162 from 2001, and 109 from 2002, with the bulk of
the remaining time spent on the observations described in
\S\ref{sec:absphase}--\S\ref{sec:vela}, as well as periodic
observations to determine the pointing model of the telescope.  As
discussed in \S\ref{sec:onaxisleak}, observations to determine the
on-axis leakages were repeated at three epochs to check the stability
of the instrumental polarization. Observations to determine the
cross-polar phase offset discussed in \S\ref{sec:absphase} were
repeated twice to check for stability.  Likewise observations of the
Moon were repeated on three separate occasions during this time
period.

\begin{figure*}[t]
\epsscale{1.3}
\plotone{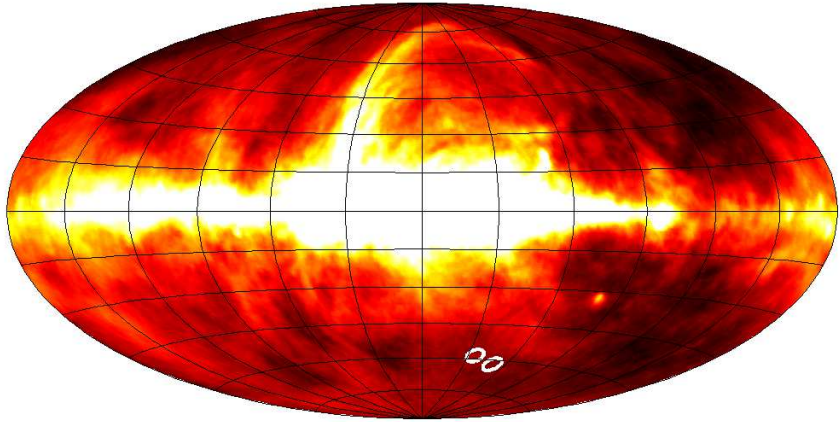}
\caption{The 408 MHz synchrotron map \markcite{haslam81}({Haslam} {et~al.} 1981) with the location of the
DASI polarization fields marked by white circles.
The color scale is linear and has been clipped at one tenth of the maximum
to allow the structure at high Galactic latitudes to be seen.}
\label{fig:haslam}
\end{figure*}

\subsection{Data Reduction}

As described in Paper I, raw visibility data from the correlators are
accumulated in 8.4-s integrations, along with monitoring data from
various hardware systems throughout the telescope.  After editing and
calibration, the raw visibilities for each 24-hour period are
combined, and these daily averages are the inputs to the analysis
presented in \polres.  Edits applied to the data can be broadly
divided into hardware and calibration edits applied to the raw
integrations, and edits applied to the 24-hour averages, and we
describe each of these in turn.

\subsubsection{Calibration and Raw Visibility Cuts}

As the first level of editing, the 8.4-s integrations are rejected
during periods when a receiver LO was out of lock, when a receiver was
warm, or when the total power for a given receiver was outside the
normal range.  These cuts collectively reject $\sim11\%$ of the data.

The only cut based on the visibility data values corresponds to a $>
30\sigma$ outlier rejection, to remove rare hardware glitches; this
cut rejects $< 0.1\%$ of the data.

The surviving 8.4-s integrations from each 24-hour period are then
calibrated for the relative amplitude and phase of the complex
multipliers on the correlator cards, determined by the periodic
injection of a broad-band noise source.  These calibration factors
have proved to be extremely repeatable in the three years that the
telescope has operated, but we nevertheless reject baselines for which
these factors lie on the tails of their distributions, indicating
possible problems with the multiplier hardware.  Data are rejected
when the relative gain of the real and imaginary multipliers falls
outside the range 0.6\dash1.2, or when the relative phase exceeds
$\pm20\deg$, regimes for which the distributions of these factors
over all baselines have become noticeably non-Gaussian.  These cuts
reject an additional $13\%$ of the data.

As described in \S\ref{sec:polresponse}, every baseline samples the four
Stokes states on a Walsh sequence.  The observations presented here
employ a period-16 Walsh sequence with a time step of 200 seconds, so
that an observation is nominally completed in 3200 seconds.  With the
inclusion of switching time for the polarizers, however, the net
observing time is somewhat less, and a reduction of $> 200~$s is
generally an indication of a polarizer sticking (see the discussion of
the polarizer assembly in \S\ref{sec:polarizers}).  In combination
with the cuts above, we therefore reject any scan with total time in a
valid Stokes state $< 3000~$s, rejecting $0.1\%$ of the data.
Calibrator scans are also rejected if previous edits have reduced the
total integration time by more than a factor of two.  Loss of a
calibrator scan results in the rejection of an entire day of data.

The data are next calibrated to remove complex instrumental gains, and
to convert to absolute flux scale, using the bracketing observations
of RCW38.  This gain calibration is equivalent to the combination of
the amplitude and phase calibration described in \paperone, but here
we construct the cross-polar gains from antenna-based fits to the
co-polar visibilities, as described in \S\ref{sec:calib}.  The flux
scale of RCW38 has been previously determined to $3.5\%$ from
measurements of absolute loads transferred to the source in 2000 and
2001, as discussed in Paper I.

All data which lack bracketing calibrator scans are rejected, as are
co-polar data for which the calibrator amplitude varies by $> 10\%$
over 24 hours, or for which the calibrator phase varies by $> 15\deg$.
For the cross-polar data, these limits are relaxed to $15\%$ and
$20\deg$, as the lower \snr\ cross-polar data are dominated by thermal
noise, and are therefore relatively insensitive to random errors in
the daily calibration.  These cuts reject $5\%$ of the data.

\subsubsection{Calibrated Visibility Cuts}
\label{sec:calcuts}

After the raw data are edited and calibrated, the 8.4-s integrations
are combined to form 24-hour averages of the visibilities for every
baseline.  (Recall that for a co-planar array, the projected baseline
lengths do not change as a source is tracked across the sky, and that
sources at the South Pole do not rotate in parallactic angle.  The
combination of visibilities on arbitrary timescales therefore
engenders no loss of information due to smearing in the Fourier
plane.)  Cuts applied to the 24-hour averages include astronomical
cuts (based on the position of the Sun and Moon), weather cuts, and
cuts derived from the visibility noise.

In the first category, co-polar visibilities are cut whenever the Sun
was above the horizon.  The cross-polar visibilities show little
evidence for contamination from the Sun at low elevations, and are
rejected only when the Sun elevation exceeds $> 5\deg$, or the Sun
is closer than $90\deg$ to the fields and above the horizon.

The Moon rises and sets once a month at the Pole, and passes within
$45\deg$ of the CMB fields. Although the data show little evidence of
contamination for Moon elevations $< 10\deg$, co-polar data were
rejected when the Moon was above the horizon and closer than $80\deg$
and $60\deg$, respectively, to the CMB and calibrator positions. A
less stringent cut is imposed on the cross-polar data, as they are
dominated by thermal noise and are therefore less sensitive to daily
calibration uncertainties.  For the 2002 data, there is evidence that
the sunshield may actually increase the data scatter through secondary
reflections when the Moon would otherwise be below the ground shield,
and the 2002 co-polar data are rejected whenever the Moon is above the
horizon.

The correlation matrix of visibilities was computed for each day of
data, over all 78 complex baselines, 10 correlator frequencies, and 4
Stokes states, where concurrent data from different Stokes states are
available (see \S\ref{sec:polresponse}).  This $6240 \times 6240$
matrix is inspected for off-diagonal correlations, and we reject
entire days when the significance of any correlation exceeds
$8\sigma$, presumably due to weather.  This cut rejects 22 days of
data not already rejected by the lunar and solar cuts ($8.1\%$ of the
data).

As an additional test of the correlator hardware, we form the sum of
visibilities over consecutive pairs of 1-h observations, averaged over
24 hours, and over all baselines for each of DASI's 10 correlators.
(Note, each correlator performs the 78 real and imaginary correlations
for one of the ten 1~GHz-wide DASI frequency bands \markcite{padin00}(see {Padin} {et~al.} 2001, and
\paperone).)  This quantity is sensitive to output of the
correlator that is coherent between observations of the two CMB
fields.  All data are rejected for an entire correlator on days when
there is evidence for offsets large compared to the thermal noise.
Not surprisingly, this statistic rejects data for all correlators
taken during sunset 2002, but also rejects data from a single
correlator which showed unusually large offsets throughout much of
2002.  Of the data not previously rejected from the astronomical and
weather cuts described above, this edit rejects an additional $1.9\%$.

Data are also rejected from an entire correlator on days when the
1-hour variance, averaged over all baselines of that correlator, is
grossly discrepant with the mean variance over 2001\dash2002.  Data are
rejected on a per-baseline basis if the variance within any 1-hour
period falls on the extreme tails of the expected reduced $\chi^2$
distribution.  Collectively, these edits reject a negligible fraction
of the data.

For all of the cuts described above, the results are insensitive to
small variations in the chosen cut levels.  As discussed in \polres,
as these cut levels are varied, evidence for residual contamination
eventually appears first in $\chi^2$ consistency tests on differenced
data sets, then as excess noise in the 1-hour observations, and only
lastly in the likelihood results.  The likelihood results are found to
be insensitive even to quite large variations in the precise edit
thresholds, and we are confident that the thresholds settings
introduce no significant bias in the CMB analysis.

\begin{figure*}[t]
\begin{center}
\epsfig{file=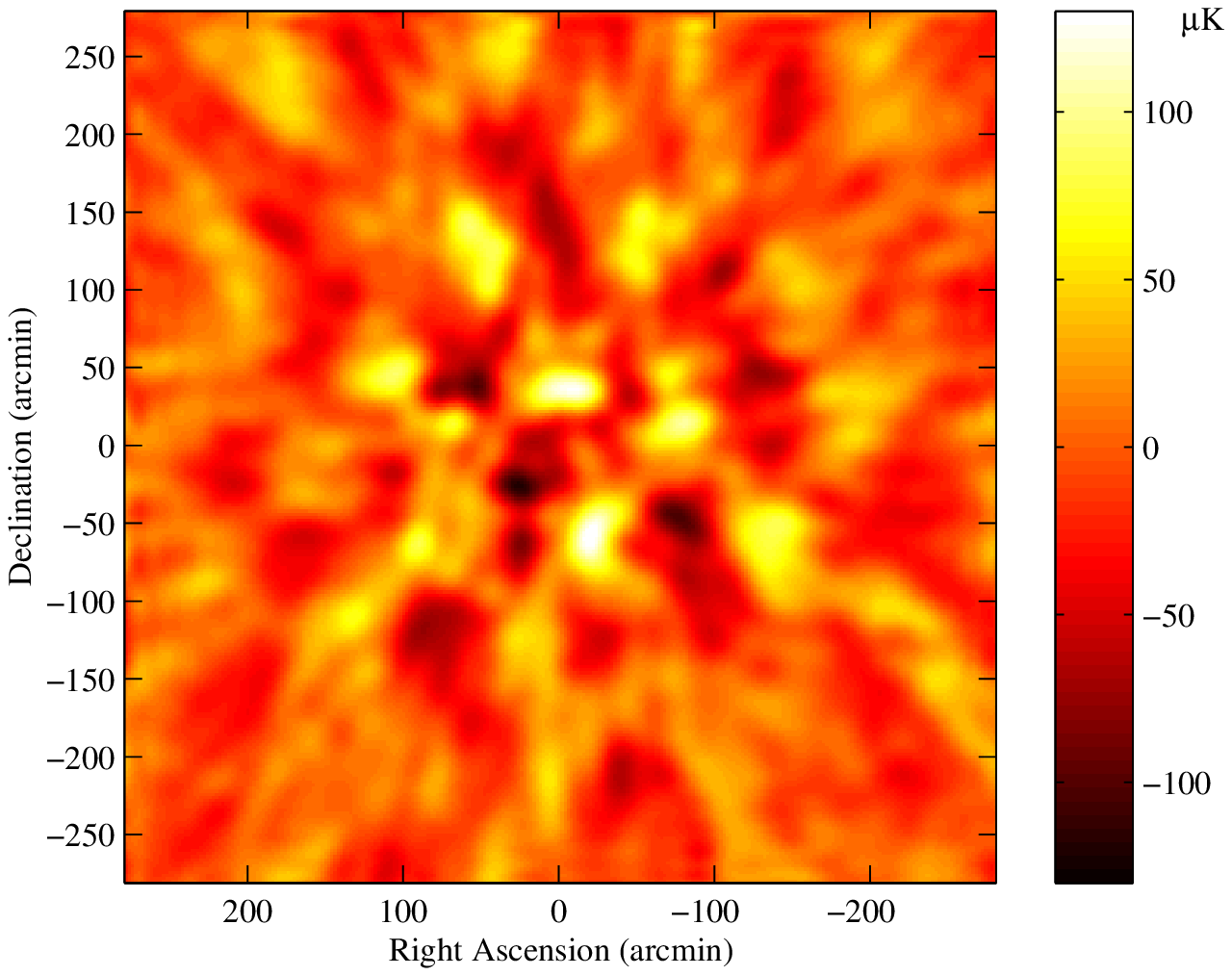,width=3in}
\epsfig{file=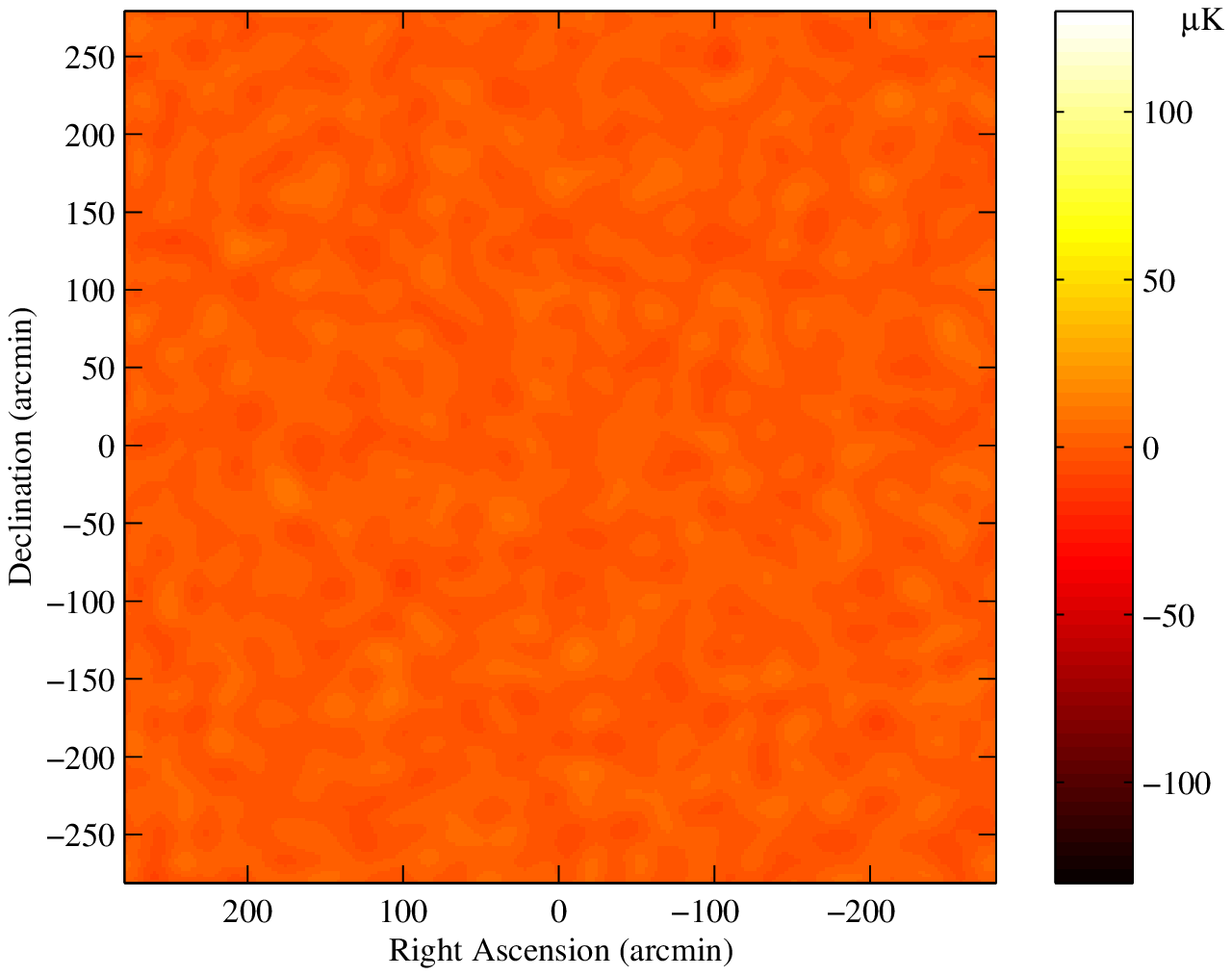,width=3in}
\end{center}
\caption{Total intensity maps of the differenced (C2--C3) CMB fields.
(left panel) Sum map of two epochs containing equal amounts of data.
(right panel) Difference map of the two epochs, plotted to the same
temperature scale.  The noise level in the map is $2.8\muK$, for a \snr\
at beam center of $\sim 25$.}
\label{fig:cmbmaps} 
\end{figure*}

\subsection{CMB Maps and Discussion}

In this experiment, two CMB fields were observed to provide a
constraint on any common signal not removed by the ground shield;
accordingly, the analysis presented in \polres~pertain to the {\it
differenced} visibilities of the two CMB fields.  As described in that
paper, the differenced visibilities are subjected to numerous
consistency tests and no evidence is found for excess noise in any of
the various subdivisions of the data by epoch, deck rotation or as a
function of baseline length.

Shown in Figure~\ref{fig:cmbmaps} are total intensity maps of the
differenced visibilities, where the maps are the sum and difference of
two epochs with equal amounts of data.  The sum map clearly shows
structure enveloped by the primary beam of the feed horns, and as
expected from the consistency tests, no residual signal is seen in the
difference map above the noise.  The detected structure is consistent
with visibilities measured for these fields in the 2000 data presented
in \paperone.  With a noise floor in the difference map of $2.8\muK$,
we achieve a \snr\ ratio of approximately $25$ at beam center, for a
synthesized beam of $\sim 22'$.

In \paperone, we gave estimates of the total intensity of synchrotron,
free-free and thermal dust emission in the region of our fields,
showing that the expected amplitudes are very small.  This was
confirmed in \papertwo by template-based cross-correlation analysis
which showed that the contribution of each of these foregrounds to our
temperature anisotropy data is negligible.  

The expected fractional polarization of the CMB is of order 10\%,
while the corresponding number for free-free emission is less than 1\%.
Diffuse thermal dust emission may be polarized at the several percent
level \markcite{hildebrand00}see, e.g., {Hildebrand} {et~al.} (2000), although the admixture of
dust and free-free emission observed with DASI in \ngc\ is
$\ll1\%$~(see \S\ref{sec:ngc6334}, and Figure~\ref{fig:ngc6334}).
Likewise, emission from spinning dust is not expected to be polarized
at detectable levels \markcite{lazarian02}(Lazarian \& Prunet 2002).  Therefore if free-free and
dust emission do not contribute significantly to the temperature
anisotropy results they are not expected to contribute to the
polarization.

As described in \S\ref{sec:field_select}, no point sources are
detected in either field, and there is at best scant evidence for a
correlation between the locations of the 31 brightest PMN sources and
the pixel values in the map.  Nevertheless these sources, whose 5~GHz
flux when enveloped with the DASI primary beam exceeds $50~$mJy, are
projected out of the co-polar data for all of the likelihood results
presented in \polres.

As discussed in \polres, with these sources projected out of the
differenced co-polar data, the temperature spectral index is
consistent with a thermal spectrum ($\beta = -0.01\pm0.15$, where
$T\propto\nu^\beta$).  Any significant non-thermal emission can be
excluded with high confidence. If none of the PMN sources are
projected out, the spectral index limit shifts to $-0.12\pm0.13$,
indicating that the PMN sources may be detected statistically, but
that the CMB is nevertheless the dominant signal in these fields.
With no PMN sources projected out of the cross-polar data, the
spectral index of the polarized visibilities is also consistent with a
thermal spectrum, though with considerably lower \snr\, $\beta =
0.17\pm1.8$ ($68\%$ confidence).  Though this result alone is not
conclusive, we note that it is inconsistent with synchrotron at nearly
the $2\sigma$ level.

\section{Conclusion}

We have described an experiment to measure polarization in the CMB,
using the Degree Angular Scale Interferometer (DASI).  New broadband
polarizers were installed in the telescope in 2001, and careful
optimization of these polarizers prior to installation resulted in
achromatic performance across DASI's frequency band to $\lesssim1\%$,
as confirmed by astronomical observations.  We have developed
techniques which allow us to calibrate the polarimeter end-to-end
using only unpolarized sources, and an extensive campaign of
observation has resulted in a characterization of the instrumental
polarization response to levels well below what is required for a
detection of CMB polarization.  Observations of the Moon and various
Galactic sources demonstrate that DASI can map degree-scale
polarization to high accuracy.

During 2001\dash2002, the telescope acquired 271 days of data in all four
Stokes parameters on two fields identified from previous observations
with DASI as containing no detectable point sources.  The data show no
evidence for contamination by point sources, polarized or unpolarized,
and both the total intensity and polarization data are consistent with
a thermal spectrum.  These observations show structure from the CMB
detected with an unprecedented \snr\ of $\sim25$.

\acknowledgments We are indebted to the Caltech CBI team led by Tony
Readhead, in particular to Steve Padin for a considerable fraction of
DASI's electronics design, to John Cartwright for the downconverter
design and to Martin Shepherd for data acquisition software. We thank
Raytheon Polar Services for their support of the DASI project,
including the erection of the DASI ground shields, and in particular
Roger Rowatt and his crew. We are indebted to the Center for
Astrophysical Research in Antarctica (CARA), in particular the efforts
of Allan Day, Stephan Meyer, Nancy Odalen, Bob, Dave and Ed Pernic, Bob
Lowenstein, Bob Spotz, Michael Whitehead and the CARA polar operations staff. We thank
Jacob Kooi for his assistance with the installation of the
polarization hardware, John Yamasaki for help with electronics, Ellen LaRue and Gene Drag for the assembly of
the calibration wire grids, Mike Loh for assistance with the sunshield
and Kim Coble for analysis of DASI primary beam measurements. We thank
the TopHat collaboration for the use of their sun shield.  We thank
the observatory staff of the Australia Telescope Compact Array, in
particular Bob Sault and Ravi Subrahmanyan, for their generosity in
providing point source observations of the DASI fields.  This research
was initially supported by the National Science Foundation (NSF) under
a cooperative agreement (OPP 89-20223) with CARA, a NSF Science and
Technology Center; it is currently supported by NSF grant
OPP-0094541. JEC gratefully acknowledges support from the James S.\
McDonnell Foundation and the David and Lucile Packard Foundation.  JEC
and CP gratefully acknowledge support from the Center for Cosmological
Physics.

\bibliography{}
\end{document}